# Using a Heterodyne Detection Scheme in a Subcarrier Wave Quantum Communication System

K.S. Melnik[1], N.M. Arslanov[1], O.I. Bannik[1], L.R. Gilyazov[1],
V.I. Egorov[2], A.V. Gleim[1,2], S.A. Moiseev[1,*]

[1] Kazan Quantum Center, Kazan National Research Technical University n.a. A.N.Tupolev-KAI,
10 K. Marx, Kazan, 420111, Russia
[2] ITMO University, St. Petersburg, 197101, Russia
*e-mail: samoi@yandex.ru

The single photon detectors currently used in quantum communication schemes impose considerable restrictions on signal registration and dark count rates, require cooling to low temperatures, and are relatively expensive. Alternative approaches have recently been proposed that are based on the homogeneous and heterogeneous detection of quantum signals and can be used with conventional photodetectors. This work studies the possibility of obtaining a heterogeneous detection scheme in a subcarrier wave (also known as homodyne, self-homodyne or self-heterodyne detection methods for side band frequencies) quantum communication system that could be used to create quantum networks.

## INTRODUCTION

Quantum networks [1] designed for transferring quantum information to large numbers of users in order to organize safe communication or data exchange between quantum computing machines are now being developed. The typical speeds of quantum bit generation in quantum networks remain relatively slow (no greater than 1–2 Mbit · s$^{-1}$) [2]. This is mainly due to the low efficiency of detection blocks. Most quantum communication schemes include different kinds of photon detectors for registering quantum states. Photon detectors based on avalanche-type light-emitting diodes are characterized by relatively high levels of dark operations (around 1–10 kHz), and in practical terms have limited registering speeds, due to the need to compensate for the effect of afterpulsing. Superconducting detectors [3] have high quantum efficiency (up to 80% for a wavelength of 1.5 μm), low jitter (100 ps), slow dark counts (around 10 Hz), and are able to determine the number of photons in a count [4]. To ensure their functioning, however, we must cool the sensitive element to temperatures of ~2.5 K. In addition, using devices for counting single photons in quantum communication systems imposes substantial restrictions on their application upon the spectral multiplexing of a quantum channel together with classic information channels.

In addition to schemes that use counts of single photons for registering single-photon qubits, there is an alternative approach known as the system of quantum communication on continuous variables, which has been under development for the last decade [5, 6]. Here, the quantum states of photon qubits is measured via homodyne detection by mixing the local oscillator field with signal radiation, which carries information about the state of photon qubits. An advantage of this approach is high count speed and the simple design of the detector.

The authors of [7] proposed the scheme of quantum communication on side frequencies (QCSF). A distinguishing feature of the QCSF system compared to other schemes is its way of generating and encoding single-photon states. Single photons are not irradiated directly by a source, but are carried to the side frequencies of spectrum as a result of phase modulation by the radio frequency signal of intense single-frequency laser irradiation. At the same time, the phase of the modulating signal determines the phase shift between radiation on the central and side frequencies, allowing us to implement protocols of quantum communication B92 [8], BB84 [9], and so on. Due to the selected shape of the optical signal scheme, QCSF has considerable advantages in creating quantum networks based on the existing infrastructure of optical communication. These include stability against the effects external conditions have on the parameters of signal photons, unidirectionality of the optical scheme, and higher spectral efficiency [10].

This work proposes an optical scheme QCSF with heterodyne signal detection, which in the future will allow us to build a protocol of quantum communication on side frequencies by encoding signal qubits on continuous variables in the basis of coherent states.

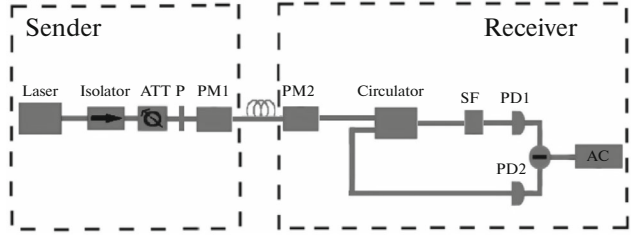

**Fig. 1.** Scheme of our system for quantum communication on side frequencies with heterodyne detection: PM1 and PM2 are the phase modulators, SF is the spectral filter, P is the polarizer, ATT is the controllable optic attenuator, PD1 and PD2 are the photodetectors, and SA is the spectrum analyzer.

## SCHEME OF QUANTUM COMMUNICATION ON SIDE FREQUENCIES WITH HETERODYNE DETECTION

Figure 1 shows the proposed scheme of a device for quantum communication on side frequencies with heterodyne signal detection. The unit consists of sender and receiver blocks. The sender block includes a semiconducting laser, isolator, polarizer and phase modulator. The receiver block is equipped with a phase modulator, circulator, spectral filter, and optical heterodyne detection scheme. The semiconducting laser generates monochromatic radiation that is sent to an isolator for absorbing reverse reflections from optical components of the device. Other radiation is directed to polarizer P, which ensures agreement between the polarization of signal radiation and the axis of the crystal of phase modulator PM1. In PM1, optical irradiation is modulated by the radio frequency signal; at the same time, additional phase shift $\varphi_1$ is introduced, which determines the information sent in signal. For modulation index $m \ll 1$, we have

$$\begin{aligned} E_1(t) &= E_0 \exp[i\omega_0 t + im\cos(\Omega t + \varphi_1)] \\ &\approx E_0' \exp(i\omega_0 t) + \frac{im}{2} E_0 \left[\exp i((\omega_0 + \Omega)t + \varphi_1)) + \exp i((\omega_0 - \Omega)t - \varphi_1)\right], \end{aligned} \quad (1)$$

where $\omega_0$ is the optical carrier frequency, $\Omega$ is the frequency of the radio frequency signal,

$$E_0' = E_0 J_0(m)|_{m \ll 1} \cong E_0,$$

and $J_0(m)$ is a zero order Bessel function.

The modulation index set by the amplitude of the electric field supplied to the modulator is selected so as to ensure the field intensity on the side frequencies ($\omega_0 \pm \Omega$) that corresponds to the required level of signal. The parameters of the signal field are determined by the requirements of the protocol of quantum communication. Encoding in a QCSF system with homodyne detection can be done via quadrature modulation of the radio frequency control signal, and is also determined by a protocol [9].

Radiation (1) with the resulting spectrum and phase coding is directed by an optical communication line into the receiver block, where it is modulated again, but with high values of the modulation index. The modulation index is selected so as to ensure the equality of radiation amplitudes on central and side frequencies; this differs from the schemes of quantum communication on side frequencies proposed earlier that were based on counting single photons using a low modulation index in receiver block [9]. The central and side components of the radiation then undergo spectral separation. The signal is sent to an optical circulator, after which it is directed to a spectral filter (SF). The SF radiation on side frequencies that is transmitted arrives at photodetector PD1, while the radiation with the central frequency is reflected from the SF and sent to the second arm of the circulator (see Fig. 1), where it is registered on photodetector PD2. The difference between the two photocurrents registered on photodetectors PD1 and PD2 is measured at the output of the balance detector.

In this scheme, the radiation at the central frequency acts as a local oscillator, the field of which is mixed with signal radiation via the partial transfer of energy from the central to side frequency during its repeated modulation in the receiver block. The high field modulation index in the receiver block allows registration of the signal radiation encoded according to phase in the balance scheme and propagating on the side frequencies. At the same time, the intensity of the current at the output of the heterodyne scheme is proportional to the product of amplitudes of the weak signal of the side frequency and the powerful reference radiation of the central frequency, transferred to the side frequencies after the second modulation according to principles of currently accepted phase detection scheme.

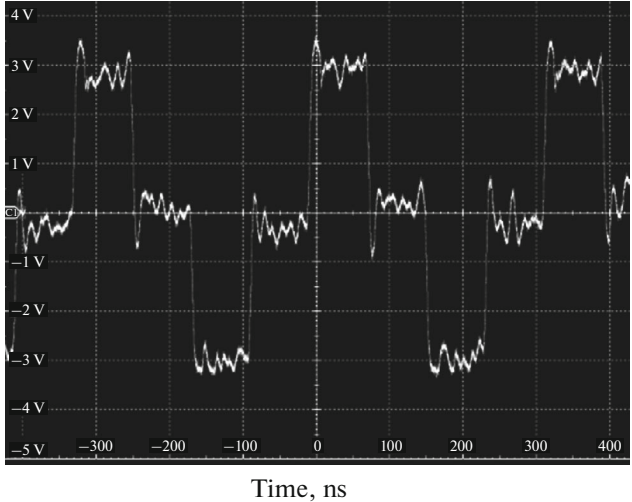

**Fig. 2.** Oscillogram showing the heterodyne registration of a signal with phase coding in our system for quantum communication on side frequencies.

An advantage of the proposed scheme of quantum communication with heterodyne detection over existing schemes of phase detection is the simplification of local oscillator signal transfer from the sender block to the receiver block so that its coherency is maintained with the variable (information) component of the field on the side frequencies. This is achieved by forming quantum signals through modulation of the local oscillator radiation and minor spectral detuning of the side frequencies with respect to the signal at the central frequency (no greater than 2.5–5 GHz [8, 9]). With such minor detuning, distortions of the spectral components of a signal in the optical tract are virtually identical. There is thus no need for a system to compensate for negative impacts on the communication line, which is required in devices with time division of the local oscillator and quantum signals. The considered approach to the phase detection of a signal greatly simplifies the optical scheme and lifts design restrictions on the limiting speed of transmitting quantum states.

## EXPERIMENTAL

The functioning of our scheme was confirmed in the regime of the classical (intense) registering of signal radiation formed on side frequencies. The signal was formed by a DFB-laser with a wavelength of 1550.12 nm and a bandwidth of 1 MHz; the emission power was 20 mW. The signal was modulated by phase electrooptical modulators with a bandwidth of 10 GHz. The modulating signal was formed by a phase-lock-loop frequency control (PLLFC) device and had a carrier frequency of 4.8 GHz. The encoding of quantum bit states was modeled via the sequential switching of the phase of the modulating 12.5 MHz radio frequency signal in the receiver block between four basic states: [0; $\pi/2$; $\pi$; $3\pi/2$]. The phase of the radio frequency signal was recorded. Switch commands were sent from a built-in card based on FPDL. The optical power of the signal at the central frequency on the output of sender modulator was 600 μW; on the side frequencies, it totalled 500 nW. The radiation on the central and side frequencies was separated by a spectral filter based on an optic fiber Bragg grating with an extinction coefficient of 30 dB. The process of generation of quantum signals was controlled by specialized software in NauLinux.

The signal was registered by a balance detector containing two PIN photodiode modules with dark current of 0.03–0.16 nA, an operating bandwidth of 2 GHz, and a sensitivity of 0.9 A W$^{-1}$. Amplification was performed by an transimpedance amplifier with a feedback resistance of 10 kΩ, and a sequentially connected voltage amplifier with an amplification coefficient of 10 dB. Figure 2 shows an oscillogram of a signal obtained at the detector output.

We can see that the proposed scheme allowed registration of the states of destructive and constructive interference on the side frequencies in the heterodyne scheme of the QCSF. The constructive interference corresponds to a voltage level of 3.5 V; destructive interference, to a voltage level of 3.2 V, which corresponds to a photocurrent upstream of the amplifier of 35 μA for constructive interference and 32 μA for destructive interference. These values corresponds to ratio between the variable component (depending on the phase shift of the radiation on side frequencies with respect to the radiation at the central frequency) of the optical signal registered at the output of the balance detector and a signal of 18 dB on the side frequencies at the output of the sender modulator. The proposed method thus allows registration of the phase of a weak optical signal by means of heterodyning.

## CONCLUSIONS

An optical scheme for a device of quantum communication on side frequencies with heterodyne detection was proposed for the first time. The possibility of registering the phase of weak signals on side frequencies using radiation transferred by a sender at the central frequency as a local oscillator was confirmed experimentally. By slightly reducing the index of modulation and the self-noise of the balance detector, this scheme has the potential for registering the quantum states of light fields transferred on side frequencies. These results can be used to create devices for quantum communications based on continuous variables with the encoding of qubits on the states of light fields with side frequencies. Compared to existing schemes, the proposed approach would help solve the problem of transmitting a local oscillator field long distances over fiber by preserving of its coherency with the variable component of the signal modulated according to phase. This would greatly simplify the design of the detector block while maintaining the high speed and distance of sending quantum bits.

ACKNOWLEDGMENTS

This work was supported by the RF Ministry of Education and Science, project nos. 03.G25.31.0229 (V.I. Egorov and A.V. Gleim), 3.6982.2017/9.10 (O.I. Bannik), 3.6583.2017/9.10 (L.R. Gilyazov), and 3.7089.2017/9.10 (N.M. Arslanov).